# Sedeonic relativistic quantum mechanics

Victor L. Mironov and Sergey V. Mironov

Institute for physics of microstructures RAS, 603950, Nizhniy Novgorod, GSP-105, Russia

E-mail: mironov@ipm.sci-nnov.ru

We represent sixteen-component values "sedeons", generating associative noncommutative space-time algebra. We demonstrate a generalization of relativistic quantum mechanics using sedeonic wave functions and sedeonic space-time operators. It is shown that the sedeonic second-order equation for the sedeonic wave function, obtained from the Einstein relation for energy and momentum, describes particles with spin 1/2. We show that for the special types of wave functions the sedeonic second-order equation can be reduced to the set of sedeonic first-order equations analogous to the Dirac equation. At the same time it is shown that these sedeonic equations differ in space-time properties and describe several types of massive and corresponding massless particles. In particular we proposed four different equations, which could describe four types of neutrinos.

## 1. Introduction

It is known that scalar Schrödinger and Klein-Gordon equations for scalar wave function do not describe spin properties of quantum particles [1,2]. For the spin description W.Pauli and P.A.M.Dirac proposed matrix equations for the multicomponent spinor wave functions [3,4]. In the latter years many authors considered alternative possibilities to describe quantum particles by multicomponent wave functions on the basis of various systems of hypercomplex numbers [5-23]. The simplest generalizations of quantum mechanics based on quaternionic wave functions with spatial structure enclosing scalar and vector components were made in Refs. 5 - 11. However the essential imperfection of the quaternionic algebra is that the quaternions do not include pseudoscalar and pseudovector components. The consideration of total symmetry with respect to spatial inversion leads us to the eight-component wave functions enclosing scalar, pseudoscalar, vector and pseudovector components. However attempts to describe relativistic particles by means of different eight-component hypernumbers such as biquaternions [4,14,15], octonions [16-20] and multivectors generating associative Clifford algebras [21-23] have not made appreciable progress. In particularly, the few attempts to describe relativistic particles by means of octonion wave functions are confronted by difficulties connected with octonions nonassociativity [18]. Moreover all systems of hypercomplex numbers, which have been applied up to now for the generalization of quantum mechanics (quaternions, biquaternions, octonions and multivectors) are the objects of hypercomplex space and do not have any consistent space-geometric interpretation. Recently we proposed eight-component values "octons" [24-28] generating a closed noncommutative associative algebra and having a clear well-defined geometric interpretation. It was shown that equations of relativistic quantum mechanics can be adequately generalized on the basis of octonic wave functions and octonic spatial operators. However all above mentioned eight-component wave functions do not describe the properties of quantum particles concerned with time transformation. The consideration of total space-time symmetry requires sixteen-component wave functions.

There are some publications describing the attempts to develop quantum mechanics using different sixteen-component hypernumbers. In particular, one of approaches is the application of hypernumbers sedenions, which are obtained from octonions by Cayley-Dickson extension procedure [29-32]. But as in the case of octonions the essential imperfection of sedenions is their nonassociativity. Another approach is the description of quantum particles by hypercomplex multivectors generating associative space-time Clifford algebras. The basic idea of such multivectors is an introduction of additional noncommutative time unit, which is orthogonal to the space units [33,34]. However the application of such multivectors in quantum mechanics is considered in general as one of abstract algebraic scheme enables the reformulation of Dirac equation for the multicomponent wave functions but does not touch the physical entity of this equation.

In this paper we represent sixteen-component values "sedeons", which are the generalization of the proposed previously "octons" and generate associative noncommutative space-time algebra. On the basis of sedeonic wave functions and sedeonic space-time operators the generalized equations of relativistic quantum mechanics are formulated. We show that sedeonic second-order and first-order equations differing in space-time properties enable the consideration of several types of massive and corresponding massless particles.

## 2. Algebra of sedeons

Let us consider four groups of values, which are differed with respect to spatial and time inversion.

- Absolute scalars ($a$) and absolute vectors ($\vec{A}$) are not transformed under spatial and time inversion.
- Space scalars ($b_r$) and space vectors ($\vec{B}_r$) are changed (in sign) under spatial inversion and are not transformed under time inversion.
- Time scalars ($c_t$) and time vectors ($\vec{C}_t$) are changed under time inversion and are not transformed under spatial inversion.
- Space-time scalars ($d_{rt}$) and space-time vectors ($\vec{D}_{rt}$) are changed under spatial and time inversion.

Here indexes *r* and *t* indicate the transformations (*r* for spatial inversion and *t* for time inversion), which change the corresponding values. Let us formally define the operation of spatial inversion ($R_r$) and time inversion ($R_t$), which change the sign of corresponding values:

$$\boldsymbol{R_r}: \qquad a, b_r, c_t, d_{rt}, \vec{A}, \vec{B}_r, \vec{C}_t, \vec{D}_{rt} \;\Rightarrow\; a, -b_r, c_t, -d_{rt}, \vec{A}, -\vec{B}_r, \vec{C}_t, -\vec{D}_{rt} \tag{1}$$

$$\boldsymbol{R_t}: \qquad a, b_r, c_t, d_{rt}, \vec{A}, \vec{B}_r, \vec{C}_t, \vec{D}_{rt} \;\Rightarrow\; a, b_r, -c_t, -d_{rt}, \vec{A}, \vec{B}_r, -\vec{C}_t, -\vec{D}_{rt} \tag{2}$$

All introduced values can be integrated into one space-time object. For this purpose in the present paper we propose the special sixteen-component values, which will be named "sedeons" (in contrast to sedenions).

The sixteen-component sedeon $\tilde{\boldsymbol{S}}$ is defined by the following expression:

$$\tilde{\boldsymbol{S}} = a + \vec{A} + b_r + \vec{B}_r + c_t + \vec{C}_t + d_{rt} + \vec{D}_{rt} \,. \tag{3}$$

The sedeon (3) can be written also in the expanded form

$$\begin{aligned} \tilde{\boldsymbol{S}} &= a\boldsymbol{e} + A_1\boldsymbol{i} + A_2\boldsymbol{j} + A_3\boldsymbol{k} + b\boldsymbol{e_r} + B_1\boldsymbol{i_r} + B_2\boldsymbol{j_r} + B_3\boldsymbol{k_r} \\ &+ c\boldsymbol{e}_t + C_1\boldsymbol{i_t} + C_2\boldsymbol{j_t} + C_3\boldsymbol{k_t} + d\boldsymbol{e}_{rt} + D_1\boldsymbol{i_{rt}} + D_2\boldsymbol{j_{rt}} + D_3\boldsymbol{k_{rt}} \,, \end{aligned} \tag{4}$$

where values $\boldsymbol{i}$, $\boldsymbol{j}$ and $\boldsymbol{k}$ are absolute unit vectors; $\boldsymbol{i_r}$, $\boldsymbol{j_r}$ and $\boldsymbol{k_r}$ are space unit vectors; $\boldsymbol{i_t}$, $\boldsymbol{j_t}$ and $\boldsymbol{k_t}$ are time unit vectors; $\boldsymbol{i_{rt}}$, $\boldsymbol{j_{rt}}$ and $\boldsymbol{k_{rt}}$ are space-time unit vectors; $\boldsymbol{e}$ is absolute scalar unit ($\boldsymbol{e} \equiv 1$); $\boldsymbol{e_r}$ is space scalar unit; $\boldsymbol{e_t}$ is time scalar unit; $\boldsymbol{e_{rt}}$ is space-time scalar unit. Let $\boldsymbol{i}, \boldsymbol{j}, \boldsymbol{k}$; $\boldsymbol{i_r}, \boldsymbol{j_r}, \boldsymbol{k_r}$; $\boldsymbol{i_t}, \boldsymbol{j_t}, \boldsymbol{k_t}$ and $\boldsymbol{i_{rt}}, \boldsymbol{j_{rt}}, \boldsymbol{k_{rt}}$ be the right Cartesian bases and corresponding unit vectors are parallel to each other. The sedeonic components

$$a,\ b,\ c,\ d,\ A_1,\ A_2,\ A_3, B_1,\ B_2,\ B_3,\ C_1,\ C_2,\ C_3,\ D_1,\ D_2,\ D_3$$

are numbers (complex in general). The values

$$\boldsymbol{e}, \boldsymbol{i}, \ \boldsymbol{j}, \boldsymbol{k}, \boldsymbol{e}_r, \boldsymbol{i}_r, \ \boldsymbol{j}_r, \boldsymbol{k}_r, \boldsymbol{e}_t, \boldsymbol{i}_t, \ \boldsymbol{j}_t, \boldsymbol{k}_t, \boldsymbol{e}_{rt}, \boldsymbol{i}_{rt}, \ \boldsymbol{j}_{rt}, \boldsymbol{k}_{rt} \tag{5}$$

are the space-time basis of sedeon. The rules for multiplication of basis elements (5) are formulated taking into account the symmetry of their products with respect to the operations of spatial and time inversion.

The squares of sedeonic scalar units and unit vectors are positively defined and equal to 1.

$$\boldsymbol{e}^2 = \boldsymbol{i}^2 = \boldsymbol{j}^2 = \boldsymbol{k}^2 = \boldsymbol{e}_r^2 = \boldsymbol{i}_r^2 = \boldsymbol{j}_r^2 = \boldsymbol{k}_r^2 = \boldsymbol{e}_t^2 = \boldsymbol{i}_t^2 = \boldsymbol{j}_t^2 = \boldsymbol{k}_t^2 = \boldsymbol{e}_{rt}^2 = \boldsymbol{i}_{rt}^2 = \boldsymbol{j}_{rt}^2 = \boldsymbol{k}_{rt}^2 = 1. \tag{6}$$

The units $\boldsymbol{e}_r, \boldsymbol{e}_t, \boldsymbol{e}_{rt}$ commute with each other and with all sedeonic unit vectors. The unit vectors satisfy following relations

$$\boldsymbol{i}_\upsilon = \boldsymbol{i}\,\boldsymbol{e}_\upsilon, \ \boldsymbol{j}_\upsilon = \boldsymbol{j}\,\boldsymbol{e}_\upsilon, \ \boldsymbol{k}_\upsilon = \boldsymbol{k}\,\boldsymbol{e}_\upsilon, \tag{7}$$

where index $\upsilon$ takes any meaning from the set of $\{r, t, rt\}$. Besides unit vectors anticommute with each other:

$$\boldsymbol{i}\,\boldsymbol{j} = -\boldsymbol{j}\,\boldsymbol{i}; \ \boldsymbol{i}\,\boldsymbol{k} = -\boldsymbol{k}\,\boldsymbol{i}; \ \boldsymbol{k}\,\boldsymbol{j} = -\boldsymbol{j}\,\boldsymbol{k}. \tag{8}$$

The rules of commutation for the rest unit vectors are constructed in accordance with relations (7).

The rules of multiplication are constructed taking into account (6) - (8). For example the multiplication rules for the absolute unit vectors are

$$\boldsymbol{i}\,\boldsymbol{j} = \xi \boldsymbol{k}; \ \boldsymbol{j}\boldsymbol{k} = \xi \boldsymbol{i}; \ \boldsymbol{k}\,\boldsymbol{i} = \xi\, \boldsymbol{j}, \tag{9}$$

where the value $\xi$ is the imaginary unit ($\xi^2 = -1$). The full rules for multiplication of the elements of sedeonic basis are represented in the Appendix A. Actually these rules can be represented by means of two simple tables describing multiplication of absolute unit vectors $\boldsymbol{i}, \boldsymbol{j}, \boldsymbol{k}$ and sedeonic units $\boldsymbol{e}_r, \boldsymbol{e}_t, \boldsymbol{e}_{rt}$ (see tables 1 and 2).

*Table 1.*

| | $\boldsymbol{i}$ | $\boldsymbol{j}$ | $\boldsymbol{k}$ |
|---|---|---|---|
| $\boldsymbol{i}$ | 1 | $\xi\boldsymbol{k}$ | $-\xi\boldsymbol{j}$ |
| $\boldsymbol{j}$ | $-\xi\boldsymbol{k}$ | 1 | $\xi\boldsymbol{i}$ |
| $\boldsymbol{k}$ | $\xi\boldsymbol{j}$ | $-\xi\boldsymbol{i}$ | 1 |

*Table 2.*

| | $\boldsymbol{e}_r$ | $\boldsymbol{e}_t$ | $\boldsymbol{e}_{rt}$ |
|---|---|---|---|
| $\boldsymbol{e}_r$ | 1 | $\boldsymbol{e}_{rt}$ | $\boldsymbol{e}_t$ |
| $\boldsymbol{e}_t$ | $\boldsymbol{e}_{rt}$ | 1 | $\boldsymbol{e}_r$ |
| $\boldsymbol{e}_{rt}$ | $\boldsymbol{e}_t$ | $\boldsymbol{e}_r$ | 1 |

We would like to emphasize especially that sedeonic algebra is associative. The property of associativity follows directly from multiplication rules.

Thus the sedeon $\tilde{\boldsymbol{S}}$ is the complicated space-time object consisting of absolute scalar, space scalar, time scalar, space-time scalar, absolute vector, space vector, time vector and space-time vector. Note that $1, \boldsymbol{i}, \boldsymbol{j}, \boldsymbol{k}$ is distinguish sedeonic basis since the corresponding components of sedeon are not transformed under spatial and time inversion. Taking into account the relations between different elements of sedeonic basis (Appendix A) the sedeon can be represented in the compact form. Introducing new designations of sedeon-scalars

$$\boldsymbol{W}_0 = (a + b\boldsymbol{e}_r + c\boldsymbol{e}_t + d\boldsymbol{e}_{rt}),$$

$$\boldsymbol{W}_1 = (A_1 + B_1\boldsymbol{e}_r + C_1\boldsymbol{e}_t + D_1\boldsymbol{e}_{rt}),$$

$$\boldsymbol{W}_2 = (A_2 + B_2\boldsymbol{e}_r + C_2\boldsymbol{e}_t + D_2\boldsymbol{e}_{rt}), \qquad (10)$$
$$\boldsymbol{W}_3 = (A_3 + B_3\boldsymbol{e}_r + C_3\boldsymbol{e}_t + D_3\boldsymbol{e}_{rt}),$$

we can write the sedeon (4) as

$$\tilde{\boldsymbol{S}} = \boldsymbol{W}_0 + \boldsymbol{W}_1\boldsymbol{i} + \boldsymbol{W}_2\boldsymbol{j} + \boldsymbol{W}_3\boldsymbol{k} \qquad (11)$$

or introducing the sedeon-vector

$$\vec{\boldsymbol{W}} = \boldsymbol{W}_1\boldsymbol{i} + \boldsymbol{W}_2\boldsymbol{j} + \boldsymbol{W}_3\boldsymbol{k} \qquad (12)$$

the sedeon can be represented in a very compact form

$$\tilde{\boldsymbol{S}} = \boldsymbol{W}_0 + \vec{\boldsymbol{W}}\,. \qquad (13)$$

Further we will indicate sedeon-scalars and sedeon-vectors with the bold capital letters.

Let us consider the rules of sedeonic multiplication in detail. In correspondence with rules of multiplication for sedeon basis elements the sedeonic product of two sedeons can be represented in the following form

$$\tilde{\boldsymbol{S}}_1\tilde{\boldsymbol{S}}_2 = \left(\boldsymbol{W}_{10} + \vec{\boldsymbol{W}}_1\right)\left(\boldsymbol{W}_{20} + \vec{\boldsymbol{W}}_2\right) = \boldsymbol{W}_{10}\boldsymbol{W}_{20} + \boldsymbol{W}_{10}\vec{\boldsymbol{W}}_2 + \boldsymbol{W}_{20}\vec{\boldsymbol{W}}_1 + \left(\vec{\boldsymbol{W}}_1 \cdot \vec{\boldsymbol{W}}_2\right) + \left[\vec{\boldsymbol{W}}_1 \times \vec{\boldsymbol{W}}_2\right]. \qquad (14)$$

Here we denoted the sedeonic scalar multiplication of two sedeon-vectors (internal product) by symbol "$\cdot$" and round brackets

$$\left(\vec{\boldsymbol{W}}_1 \cdot \vec{\boldsymbol{W}}_2\right) = \boldsymbol{W}_{11}\boldsymbol{W}_{21} + \boldsymbol{W}_{12}\boldsymbol{W}_{22} + \boldsymbol{W}_{13}\boldsymbol{W}_{23}, \qquad (15)$$

and sedeonic vector multiplication (external product) by symbol "$\times$" and square brackets,

$$\left[\vec{\boldsymbol{W}}_1 \times \vec{\boldsymbol{W}}_2\right] = \xi\left(\boldsymbol{W}_{12}\boldsymbol{W}_{23} - \boldsymbol{W}_{13}\boldsymbol{W}_{22}\right)\boldsymbol{i} + \xi\left(\boldsymbol{W}_{13}\boldsymbol{W}_{21} - \boldsymbol{W}_{11}\boldsymbol{W}_{23}\right)\boldsymbol{j} + \xi\left(\boldsymbol{W}_{11}\boldsymbol{W}_{22} - \boldsymbol{W}_{12}\boldsymbol{W}_{21}\right)\boldsymbol{k}\,. \qquad (16)$$

In (15) and (16) the component multiplication is performed in accordance with the table 2. Thus the sedeonic product

$$\tilde{\boldsymbol{F}} = \tilde{\boldsymbol{S}}_1\tilde{\boldsymbol{S}}_2 = \boldsymbol{F}_0 + \vec{\boldsymbol{F}} \qquad (17)$$

has the following components:

$$\boldsymbol{F}_0 = \boldsymbol{W}_{10}\boldsymbol{W}_{20} + \boldsymbol{W}_{11}\boldsymbol{W}_{21} + \boldsymbol{W}_{12}\boldsymbol{W}_{22} + \boldsymbol{W}_{13}\boldsymbol{W}_{23},$$
$$\boldsymbol{F}_1 = \boldsymbol{W}_{10}\boldsymbol{W}_{21} + \boldsymbol{W}_{20}\boldsymbol{W}_{11} + \xi\left(\boldsymbol{W}_{12}\boldsymbol{W}_{23} - \boldsymbol{W}_{13}\boldsymbol{W}_{22}\right), \qquad (18)$$
$$\boldsymbol{F}_2 = \boldsymbol{W}_{10}\boldsymbol{W}_{22} + \boldsymbol{W}_{20}\boldsymbol{W}_{12} + \xi\left(\boldsymbol{W}_{13}\boldsymbol{W}_{21} - \boldsymbol{W}_{11}\boldsymbol{W}_{23}\right),$$
$$\boldsymbol{F}_3 = \boldsymbol{W}_{10}\boldsymbol{W}_{23} + \boldsymbol{W}_{20}\boldsymbol{W}_{13} + \xi\left(\boldsymbol{W}_{11}\boldsymbol{W}_{22} - \boldsymbol{W}_{12}\boldsymbol{W}_{21}\right).$$

In the next sections we apply the sedeonic algebra to the generalization of electrodynamics and relativistic quantum mechanics.

## 3. Sedeonic representation of electromagnetic field equations

In this section we construct generalized sedeonic equation for the electromagnetic field by analogy with Ref. 25 and consider the restrictions imposed by Maxwell equations on the space-time structure of electromagnetic potentials and fields.

Let us consider the absolute scalar potential $\Phi$ and absolute vector potential $\vec{A} = A_x\boldsymbol{i} + A_y\boldsymbol{j} + A_z\boldsymbol{k}$. Then using the differential operator $\vec{\nabla}$ in the form

$$\vec{\nabla} = \frac{\partial}{\partial x}\boldsymbol{i} + \frac{\partial}{\partial y}\boldsymbol{j} + \frac{\partial}{\partial z}\boldsymbol{k}, \qquad (19)$$

the generalized sedeonic equation of electrodynamics can be written in the following compact form:

$$\left(\frac{1}{c}\frac{\partial}{\partial t}-\alpha\vec{\nabla}\right)\left(\frac{1}{c}\frac{\partial}{\partial t}+\alpha\vec{\nabla}\right)\left(\Phi+\beta\vec{A}\right)=\gamma 4\pi\rho+\delta\frac{4\pi}{c}\vec{j}\,. \qquad (20)$$

Here indefinite coefficients $\alpha$, $\beta$, $\gamma$, $\delta$ can take meanings from the set of values $\{1,\boldsymbol{e}_r,\boldsymbol{e}_t,\boldsymbol{e}_{rt}\}$. Since potentials $\Phi$ and $\vec{A}$ should satisfy the corresponding wave equations, we obtain that $\alpha^2=1$, $\gamma=1$, $\delta=\beta$. Applying one of operators in equation (20) to the sedeon of electromagnetic potentials, we get

$$\left(\frac{1}{c}\frac{\partial}{\partial t}+\alpha\vec{\nabla}\right)\left(\Phi+\beta\vec{A}\right)=\frac{1}{c}\frac{\partial\Phi}{\partial t}+\alpha\vec{\nabla}\Phi+\beta\frac{1}{c}\frac{\partial\vec{A}}{\partial t}+\alpha\beta\left(\vec{\nabla}\cdot\vec{A}\right)+\alpha\beta\left[\vec{\nabla}\times\vec{A}\right]. \qquad (21)$$

For correct definition of electric and magnetic field we should require $\alpha=\beta$. Then electric and magnetic fields are defined in standard sedeonic form

$$\vec{E}=-\frac{1}{c}\frac{\partial\vec{A}}{\partial t}-\vec{\nabla}\Phi\;,\quad \vec{H}=-\xi\left[\vec{\nabla}\times\vec{A}\right]. \qquad (22)$$

Using the Lorentz gauge

$$\frac{1}{c}\frac{\partial\Phi}{\partial t}+\left(\vec{\nabla}\cdot\vec{A}\right)=0\,, \qquad (23)$$

we can rewrite the expression (21) in the following form:

$$\left(\frac{1}{c}\frac{\partial}{\partial t}+\alpha\vec{\nabla}\right)\left(\Phi+\alpha\vec{A}\right)=-\alpha\vec{E}+\xi\vec{H}\,. \qquad (24)$$

Then sedeonic equation (20) can be written as

$$\left(\frac{1}{c}\frac{\partial}{\partial t}-\alpha\vec{\nabla}\right)\left(\xi\vec{H}-\alpha\vec{E}\right)=4\pi\rho+\alpha\frac{4\pi}{c}\vec{j}\,. \qquad (25)$$

Applying the operator in the left part of equation (25) to the sedeon of the electromagnetic field we get

$$\xi\frac{1}{c}\frac{\partial\vec{H}}{\partial t}-\xi\alpha\left(\vec{\nabla}\cdot\vec{H}\right)-\xi\alpha\left[\vec{\nabla}\times\vec{H}\right]-\alpha\frac{1}{c}\frac{\partial\vec{E}}{\partial t}+\alpha^2\left(\vec{\nabla}\cdot\vec{E}\right)+\alpha^2\left[\vec{\nabla}\times\vec{E}\right]=4\pi\rho+\alpha\frac{4\pi}{c}\vec{j}\,. \qquad (26)$$

The value $\alpha$ can take any meaning from the set of $\{\boldsymbol{e}_r,\boldsymbol{e}_t,\boldsymbol{e}_{rt}\}$.

Separating values of different types in (26) we obtain the system of Maxwell equations in sedeonic form

$$\begin{cases}\left(\alpha\vec{\nabla}\cdot\alpha\vec{E}\right)=4\pi\rho,\\ \left[\alpha\vec{\nabla}\times\alpha\vec{E}\right]=-\dfrac{\xi}{c}\dfrac{\partial\vec{H}}{\partial t},\\ \left(\alpha\vec{\nabla}\cdot\vec{H}\right)=0,\\ \left[\alpha\vec{\nabla}\times\vec{H}\right]=\dfrac{\xi}{c}\dfrac{\partial\alpha\vec{E}}{\partial t}+\dfrac{4\pi\xi}{c}\alpha\vec{j}.\end{cases} \qquad (27)$$

The system (27) can be transformed to the following form:

$$\begin{cases} \left(\vec{\nabla}\cdot\vec{E}\right)=4\pi\rho, \\ \left[\vec{\nabla}\times\vec{E}\right]=-\dfrac{\xi}{c}\dfrac{\partial\vec{H}}{\partial t}, \\ \left(\vec{\nabla}\cdot\vec{H}\right)= 0, \\ \left[\vec{\nabla}\times\vec{H}\right]=\dfrac{\xi}{c}\dfrac{\partial\vec{E}}{\partial t}+\dfrac{4\pi\xi}{c}\vec{j}. \end{cases} \tag{28}$$

Thus as it follows from (28) the system of Maxwell equations can be formulated in the terms of absolute values.

Taking into account all above mentioned conditions we can represent the wave equation (20) in the following form:

$$\left(\frac{1}{c}\frac{\partial}{\partial t}-\alpha\vec{\nabla}\right)\left(\frac{1}{c}\frac{\partial}{\partial t}+\alpha\vec{\nabla}\right)\left(\Phi+\alpha\vec{A}\right)=4\pi\rho+\alpha\frac{4\pi}{c}\vec{j}\ . \tag{29}$$

The concrete space-time structure of this wave equation is defined by the coefficient $\alpha$ but it does not influence on the form of Maxwell equations.

Thus several types of sedeonic wave equations corresponding to the different $\alpha\in\{\pm\boldsymbol{e}_r,\pm\boldsymbol{e}_t,\pm\boldsymbol{e}_{rt}\}$ are possible. Since the electromagnetic potentials are real values the separation in equation (26) can be made on the base of complex values (real and imaginary parts), so $\alpha$ can be also equal to ± 1. Moreover for the real electromagnetic potentials the sedeon values

$$\pm\left(1-\boldsymbol{e}_r-\boldsymbol{e}_t-\boldsymbol{e}_{rt}\right), \tag{30}$$

$$\pm\left(1-\boldsymbol{e}_r+\boldsymbol{e}_t+\boldsymbol{e}_{rt}\right), \tag{31}$$

$$\pm\left(1+\boldsymbol{e}_r-\boldsymbol{e}_t+\boldsymbol{e}_{rt}\right), \tag{32}$$

$$\pm\left(1+\boldsymbol{e}_r+\boldsymbol{e}_t-\boldsymbol{e}_{rt}\right), \tag{33}$$

can be used as $\alpha$ coefficients since their squares are also equal to 1. The wave equations for potentials obtained from (29) with various $\alpha$ are differed in transformational properties but corresponding Maxwell equations for the field's intensities (28) are the same.

## 4. Sedeonic second-order equations of relativistic quantum mechanics

Previously [26, 27] we proposed the octonic relativistic second-order equation describing particles with spin 1/2. In addition the sedeon's algebra takes into account transformational properties of values with respect to time inversion. In this section by analogy with [27] we will construct sedeonic second-order equation and consider its space-time properties.

Let us consider the wave function of a relativistic particle in the form of a sixteen-component sedeon

$$\tilde{\psi}=\psi_0+\vec{\psi}=\psi_0+\psi_1\boldsymbol{i}+\psi_2\boldsymbol{j}+\psi_3\boldsymbol{k} \tag{34}$$

with components

$$\begin{aligned} \psi_0 &= (A_0+B_0\boldsymbol{e}_r+C_0\boldsymbol{e}_t+D_0\boldsymbol{e}_{rt}), \\ \psi_1 &= (A_1+B_1\boldsymbol{e}_r+C_1\boldsymbol{e}_t+D_1\boldsymbol{e}_{rt}), \\ \psi_2 &= (A_2+B_2\boldsymbol{e}_r+C_2\boldsymbol{e}_t+D_2\boldsymbol{e}_{rt}), \\ \psi_3 &= (A_3+B_3\boldsymbol{e}_r+C_3\boldsymbol{e}_t+D_3\boldsymbol{e}_{rt}). \end{aligned} \tag{35}$$

The components $A_s(\vec{r},t)$, $B_s(\vec{r},t)$, $C_s(\vec{r},t)$ and $D_s(\vec{r},t)$ ($s = 0,1,2,3$) are scalar (complex in general) functions of spatial coordinates and time.

The wave function of a free particle should satisfy an equation, which is obtained from the Einstein relation between particle energy and momentum

$$E^2 - p^2c^2 = m^2c^4 . \tag{36}$$

Since classical relation (36) is quadratic form the transformation properties of energy and momentum is unessential. Let us consider absolute operators of energy and momentum

$$\hat{E} = \xi\hbar\frac{\partial}{\partial t} \text{ and } \hat{\vec{p}} = -\xi\hbar\vec{\nabla} . \tag{37}$$

Then the wave equation obtained from (36) has the following form:

$$\left(\frac{1}{c^2}\frac{\partial^2}{\partial t^2} - \Delta\right)\tilde{\psi} = -\frac{m^2c^2}{\hbar^2}\tilde{\psi} . \tag{38}$$

Here $c$ is the velocity of light, $m$ is the mass of the particle and $\hbar$ is the Plank constant. In contrast to the scalar Klein-Gordon equation the expression (38) is sedeonic equation since it is written for the sedeonic function. It is clear that each of the $\tilde{\psi}$ components satisfies the scalar Klein-Gordon equation.

By analogy with octonic approach [27] the operator in the left part of equation (38) can be represented as the product of two operators:

$$\left(\frac{1}{c}\frac{\partial}{\partial t} - \alpha\vec{\nabla}\right)\left(\frac{1}{c}\frac{\partial}{\partial t} + \alpha\vec{\nabla}\right)\tilde{\psi} = -\frac{m^2c^2}{\hbar^2}\tilde{\psi} , \tag{39}$$

where undefined coefficients $\alpha$ can take meaning $\pm\boldsymbol{e}_r$, $\pm\boldsymbol{e}_t$ or $\pm\boldsymbol{e}_{rt}$. Here we assume that the sedeonic wave function $\tilde{\psi}$ is twice continuously differentiable, so $\left[\vec{\nabla}\times\vec{\nabla}\right]\tilde{\psi} = 0$. Following the ideas of [27] we can rewrite the equation (39) in the expanded operator form

$$\left(\frac{1}{c}\frac{\partial}{\partial t} - \frac{\partial}{\partial x}\hat{\alpha}\hat{\boldsymbol{i}} - \frac{\partial}{\partial y}\hat{\alpha}\hat{\boldsymbol{j}} - \frac{\partial}{\partial z}\hat{\alpha}\hat{\boldsymbol{k}}\right)\left(\frac{1}{c}\frac{\partial}{\partial t} + \frac{\partial}{\partial x}\hat{\alpha}\hat{\boldsymbol{i}} + \frac{\partial}{\partial y}\hat{\alpha}\hat{\boldsymbol{j}} + \frac{\partial}{\partial z}\hat{\alpha}\hat{\boldsymbol{k}}\right)\tilde{\psi}(\vec{r},t) = -\frac{m^2c^2}{\hbar^2}\tilde{\psi}(\vec{r},t) , \tag{40}$$

where the space-time operators $\hat{\boldsymbol{i}}, \hat{\boldsymbol{j}}, \hat{\boldsymbol{k}}, \hat{\alpha}$ $\left(\hat{\alpha} \in \{\pm\hat{\boldsymbol{e}}_r, \pm\hat{\boldsymbol{e}}_t, \pm\hat{\boldsymbol{e}}_{rt}\}\right)$ in the left part of equation (40) transform the space-time structure of the wave function by means of sedeonic multiplication. For example, the action of the $\hat{\boldsymbol{k}}$ operator can be represented as octonic multiplication of unit vector $\boldsymbol{k}$ and sedeon $\tilde{\psi}$:

$$\hat{\boldsymbol{k}}\tilde{\psi} = \boldsymbol{k}\tilde{\psi} = \psi_3 - \xi\psi_2\boldsymbol{i} + \xi\psi_1\boldsymbol{j} + \psi_0\boldsymbol{k} .$$

Further we will use symbolic designations $\hat{\boldsymbol{i}}, \hat{\boldsymbol{j}}, \hat{\boldsymbol{k}}, \hat{\alpha}$ in the operator part of equations but $\boldsymbol{i}$, $\boldsymbol{j}$, $\boldsymbol{k}$, $\boldsymbol{e}_r$, $\boldsymbol{e}_t$ and $\boldsymbol{e}_{rt}$ designations in the wave functions. The rules of multiplication and commutation for space-time operators are analogues to the rules for corresponding elements of sedeonic basis.

To describe a particle in an external electromagnetic field the following change of quantum-mechanical operators should be made [2]:

$$\hat{E} \to \hat{E} - e\Phi , \qquad \hat{\vec{p}} \to \hat{\vec{p}} - \frac{e}{c}\vec{A} , \tag{41}$$

where $\varPhi$ and $\vec{A}$ are absolute scalar and vector potentials of the electromagnetic field, $e$ is the particle charge ($e<0$ for the electron). The change (41) is equivalent to the following change of absolute differential operators:

$$\frac{\partial}{\partial t}\to\frac{\partial}{\partial t}+\frac{\xi e}{\hbar}\varPhi, \quad \vec{\nabla}\to\vec{\nabla}-\frac{\xi e}{\hbar c}\vec{A}. \tag{42}$$

Using substitution (42) we can write the equation (39) as

$$\left(\frac{1}{c}\frac{\partial}{\partial t}+\frac{\xi e}{\hbar c}\varPhi-\hat{\alpha}\vec{\nabla}+\frac{\xi e}{\hbar c}\hat{\alpha}\vec{A}\right)\left(\frac{1}{c}\frac{\partial}{\partial t}+\frac{\xi e}{\hbar c}\varPhi+\hat{\alpha}\vec{\nabla}-\frac{\xi e}{\hbar c}\hat{\alpha}\vec{A}\right)\tilde{\psi}=-\frac{m^2c^2}{\hbar^2}\tilde{\psi}. \tag{43}$$

The multiplication of sedeonic operators in the left part of (43) leads us to the following equation:

$$\left[\frac{1}{c^2}\frac{\partial^2}{\partial t^2}-\Delta+\frac{2\xi e}{\hbar c}\left(\left(\vec{A}\cdot\vec{\nabla}\right)+\frac{\varPhi}{c}\frac{\partial}{\partial t}\right)+\frac{m^2c^2}{\hbar^2}+\frac{e^2}{\hbar^2c^2}\left(A^2-\varPhi^2\right)\right]\tilde{\psi}-\frac{e}{\hbar c}\vec{H}\,\tilde{\psi}+\frac{\xi e}{\hbar c}\hat{\alpha}\vec{E}\,\tilde{\psi}=0. \tag{44}$$

Here we have taken into account that $\vec{E}=-\vec{\nabla}\varPhi-\frac{1}{c}\frac{\partial\vec{A}}{\partial t}$ is absolute vector of the electric field, $\vec{H}=-\xi\left[\vec{\nabla}\times\vec{A}\right]$ is absolute vector of the magnetic field. $\left(\vec{\nabla}\cdot\vec{A}\right)+\frac{1}{c}\frac{\partial\varPhi}{\partial t}=0$ is the condition of the Lorentz gauge. Note that the sedeonic equation (44) encloses the specific terms $\frac{e}{\hbar c}\vec{H}\,\tilde{\psi}$ and $\frac{\xi e}{\hbar c}\hat{\alpha}\vec{E}\,\tilde{\psi}$, where the fields $\vec{E}$ and $\vec{H}$ play the role of spatial sedeonic operators. It is seen that in the presence of electric field the second-order equation (44) essentially depends on the space-time type of operator $\hat{\alpha}$.

## 5. Relativistic particle in homogeneous magnetic field

Let us consider a relativistic particle in an external magnetic field directed along the $Z$ axis:

$$\vec{H}=B\boldsymbol{k}. \tag{45}$$

We select the vector potential in the gauge $\left(\vec{\nabla}\cdot\vec{A}\right)=0$:

$$\vec{A}=A_y\boldsymbol{j}=Bx\,\boldsymbol{j}. \tag{46}$$

Then the sedeonic equation for a relativistic particle (44) in the $X\,Y\,Z$ basis can be written as

$$\frac{1}{c^2}\frac{\partial^2\tilde{\psi}}{\partial t^2}-\Delta\tilde{\psi}+\frac{2\xi e}{\hbar c}Bx\frac{\partial\tilde{\psi}}{\partial y}+\frac{m^2c^2}{\hbar^2}\tilde{\psi}+\frac{e^2}{\hbar^2c^2}B^2x^2\tilde{\psi}-\frac{e}{\hbar c}\vec{H}\,\tilde{\psi}=0. \tag{47}$$

For the stationary state with the energy $E$ we get

$$\left[-\Delta+\frac{2\xi e}{\hbar c}Bx\frac{\partial}{\partial y}+\frac{m^2c^2}{\hbar^2}+\frac{e^2}{\hbar^2c^2}B^2x^2-\frac{e}{\hbar c}B\hat{\boldsymbol{k}}\right]\tilde{\psi}=\frac{E^2}{\hbar^2c^2}\tilde{\psi}. \tag{48}$$

This equation can be considered as the equation for the eigenvalues and eigenfunctions of the complicated operator placed in the left part. Since this operator commutes with operators $\hat{p}_y$ and $\hat{p}_z$ all of them have the general system of eigenfunctions. Therefore we will search a solution of (48) in the form

$$\tilde{\psi} = \tilde{\boldsymbol{W}}(x)\exp\left\{\frac{\xi}{\hbar}\left(p_y y + p_z z\right)\right\}, \tag{49}$$

where $p_y$ and $p_z$ are the motion integrals. Substituting (49) into (48) we get

$$\left[\frac{p_y^2}{\hbar^2} + \frac{p_z^2}{\hbar^2} - \frac{\partial^2}{\partial x^2} - \frac{2ep_y}{\hbar^2 c}Bx + \frac{m^2c^2}{\hbar^2} + \frac{e^2}{\hbar^2c^2}B^2x^2 - \frac{e}{\hbar c}B\hat{\boldsymbol{k}}\right]\tilde{\boldsymbol{W}} = \frac{E^2}{\hbar^2c^2}\tilde{\boldsymbol{W}}\ . \tag{50}$$

Note that operator in the left part of (50) commutes also with $\hat{\boldsymbol{k}}$, so we can search a solution as an eigenfunction of the sedeonic operator $\hat{\boldsymbol{k}}$.

The equation for the eigenvalues and eigenfunctions of operator $\hat{\boldsymbol{k}}$ has the following form:

$$\hat{\boldsymbol{k}}\tilde{\psi} = \lambda\tilde{\psi}\ , \tag{51}$$

where eigenvalues $\lambda$ are complex numbers in general. Performing octonic multiplication in the left part of (51) and equating components, we obtain the system of scalar equations, which is equivalent to the following system (see definitions (35)):

$$\begin{cases} \lambda^2 = 1, \\ \psi_3 = \lambda\psi_0, \\ \psi_2 = \xi\lambda\psi_1. \end{cases} \tag{52}$$

The first equation in (52) shows that $\lambda = \pm 1$. For each eigenvalue $\lambda$ there is a eight-dimensional subspace of eigenfunctions. Taking into account (52) we can choose the set of functions

$$\left(1+\boldsymbol{k}\right),\ \left(\boldsymbol{i}+\xi\boldsymbol{j}\right),\ \boldsymbol{e}_r\left(1+\boldsymbol{k}\right),\ \boldsymbol{e}_r\left(\boldsymbol{i}+\xi\boldsymbol{j}\right),\ \boldsymbol{e}_t\left(1+\boldsymbol{k}\right),\ \boldsymbol{e}_t\left(\boldsymbol{i}+\xi\boldsymbol{j}\right),\ \boldsymbol{e}_{rt}\left(1+\boldsymbol{k}\right),\ \boldsymbol{e}_{rt}\left(\boldsymbol{i}+\xi\boldsymbol{j}\right) \tag{53}$$

as the basis of the subspace corresponding to eigenvalue $\lambda = +1$, and set of functions

$$\left(1-\boldsymbol{k}\right),\ \left(\boldsymbol{i}-\xi\boldsymbol{j}\right),\ \boldsymbol{e}_r\left(1-\boldsymbol{k}\right),\ \boldsymbol{e}_r\left(\boldsymbol{i}-\xi\boldsymbol{j}\right),\ \boldsymbol{e}_t\left(1-\boldsymbol{k}\right),\ \boldsymbol{e}_t\left(\boldsymbol{i}-\xi\boldsymbol{j}\right),\ \boldsymbol{e}_{rt}\left(1-\boldsymbol{k}\right),\ \boldsymbol{e}_{rt}\left(\boldsymbol{i}-\xi\boldsymbol{j}\right) \tag{54}$$

as the basis of the subspace corresponding to $\lambda = -1$. Then arbitrary eigenfunctions of the operator $\hat{\boldsymbol{k}}$ corresponding to $\lambda = \pm 1$ can be represented in the form of linear combinations of the basis functions (53) or (54):

$$\tilde{\psi}_\lambda = \tilde{\boldsymbol{F}}_1^{(\lambda)}(\vec{r},t)\left(1+\lambda\boldsymbol{k}\right) + \tilde{\boldsymbol{F}}_2^{(\lambda)}(\vec{r},t)\left(\boldsymbol{i}+\xi\lambda\boldsymbol{j}\right), \tag{55}$$

where $\tilde{\boldsymbol{F}}_\mu^{(\lambda)}(\vec{r},t) = C_{\mu,1}^{(\lambda)}(\vec{r},t) + C_{\mu,2}^{(\lambda)}(\vec{r},t)\boldsymbol{e}_r + C_{\mu,3}^{(\lambda)}(\vec{r},t)\boldsymbol{e}_t + C_{\mu,4}^{(\lambda)}(\vec{r},t)\boldsymbol{e}_{rt}$; $C_{\mu,\nu}^{(\lambda)}(\vec{r},t)$ are arbitrary complex functions of space coordinates and time ($\lambda = \pm 1$; $\mu = 1, 2$; $\nu = 1, 2, 3, 4$).

So, we will search a solution of (50) in the form of (55). At that we obtain the following equation for the functions $C_{\mu,\nu}^{(\lambda)}(x)$:

$$\frac{\partial^2 C_{\mu,\nu}^{(\lambda)}}{\partial x^2} + \left[\left(\frac{E^2}{\hbar^2c^2} - \frac{p_z^2}{\hbar^2} - \frac{m^2c^2}{\hbar^2} + \lambda\frac{eB}{\hbar c}\right) - \left(\frac{eB}{\hbar c}\right)^2\left(x - \frac{cp_y}{eB}\right)^2\right]C_{\mu,\nu}^{(\lambda)} = 0\,. \tag{56}$$

This is a well known equation for the relativistic oscillator. On the basis of equation (56) we obtain the expression for the energy spectrum of a relativistic particle in a homogeneous magnetic field:

$$E_{n,\lambda}^2 = m^2c^4 + p_z^2c^2 + |e|B\hbar c\left(2n+1\right) - \lambda eB\hbar c\,. \tag{57}$$

This set of energies is absolutely identical to the energy spectrum obtained from the relativistic second-order equation following from the Dirac equation [2]. The expression (57) allows one to state that eigenvalue $\lambda$ of operator $\hat{\boldsymbol{k}}$ has the sense of spin projection and the second-order equation (44) correctly describes the interaction between spin 1/2 and the electromagnetic field.

If the wave function is the eigenfunction of the operator $\hat{\boldsymbol{k}}$, then some general statements about the spatial structure of the wave function can be made. In the stationary state with energy $E$ the wave function can be represented in the following form:

$$\tilde{\psi}_\lambda(\vec{r},t) = \left\{ \tilde{\boldsymbol{F}}_1^{(\lambda)}(\vec{r})\left(1+\lambda\boldsymbol{k}\right) + \tilde{\boldsymbol{F}}_2^{(\lambda)}(\vec{r})\left(\boldsymbol{i}+\xi\lambda\boldsymbol{j}\right) \right\} e^{-\xi\omega t}, \tag{58}$$

where $\omega = E/\hbar$. The wave function (58) is the space-time object, which we name sedeonic oscillator. The real and imaginary parts of the component $\left(1+\lambda\boldsymbol{k}\right)e^{-\xi\omega t}$ are a combination of an absolute vector directed parallel to the Z axis and an absolute scalar oscillating with the frequency $\omega$. Here the phase difference between oscillations of scalar and pseudovector parts equals 0 in case of $\lambda = 1$ or $\pi$ in case of $\lambda = -1$. The real and imaginary parts of the component $\left(\boldsymbol{i}+\lambda\xi\boldsymbol{j}\right)e^{-\xi\omega t}$ have the form of absolute vectors rotating in the plane perpendicular to the Z axis also with the frequency $\omega$. The direction of the rotation depends on the sign of $\lambda$. When $\lambda = +1$ a vector of angular velocity is directed along the Z axis but when $\lambda = -1$ this vector has the opposite direction. The rest components of the wave function (58) have a similar interpretation. The transformational properties of the wave function (58) are defined by sedeonic functions $\tilde{\boldsymbol{F}}_1^{(\lambda)}(\vec{r})$ and $\tilde{\boldsymbol{F}}_2^{(\lambda)}(\vec{r})$.

## 6. Sedeonic first-order equations

In this section we show that there is the special class of sedeonic wave functions, which describe particles with spin 1/2 but satisfy the first-order equations (analogous to the Dirac equation) differing in transformational properties with respect to spatial and time inversion.

Let us define the operators of spatial ($\hat{R}_r$) and time ($\hat{R}_t$) inversion of sedeonic wave function. The operator $\hat{R}_r$ changes the sign of spatial and space-time components of wave function (35):

$$\begin{aligned}
\hat{R}_r\psi_0 &= (A_0 - B_0\boldsymbol{e}_r + C_0\boldsymbol{e}_t - D_0\boldsymbol{e}_{rt}), \\
\hat{R}_r\psi_1 &= (A_1 - B_1\boldsymbol{e}_r + C_1\boldsymbol{e}_t - D_1\boldsymbol{e}_{rt}), \\
\hat{R}_r\psi_2 &= (A_2 - B_2\boldsymbol{e}_r + C_2\boldsymbol{e}_t - D_2\boldsymbol{e}_{rt}), \\
\hat{R}_r\psi_3 &= (A_3 - B_3\boldsymbol{e}_r + C_3\boldsymbol{e}_t - D_3\boldsymbol{e}_{rt}).
\end{aligned} \tag{59}$$

The operator $\hat{R}_t$ changes the sign of time and space-time components of wave function:

$$\begin{aligned}
\hat{R}_t\psi_0 &= (A_0 + B_0\boldsymbol{e}_r - C_0\boldsymbol{e}_t - D_0\boldsymbol{e}_{rt}), \\
\hat{R}_t\psi_1 &= (A_1 + B_1\boldsymbol{e}_r - C_1\boldsymbol{e}_t - D_1\boldsymbol{e}_{rt}), \\
\hat{R}_t\psi_2 &= (A_2 + B_2\boldsymbol{e}_r - C_2\boldsymbol{e}_t - D_2\boldsymbol{e}_{rt}), \\
\hat{R}_t\psi_3 &= (A_3 + B_3\boldsymbol{e}_r - C_3\boldsymbol{e}_t - D_3\boldsymbol{e}_{rt}).
\end{aligned} \tag{60}$$

We specially emphasize that the operators $\hat{R}_r$ and $\hat{R}_t$ do not transform arguments of the wave function.

From (59) and (60) it is easy to see that the operator $\hat{R}_r$ anti-commutes with $\hat{\boldsymbol{e}}_r$, $\hat{\boldsymbol{e}}_{rt}$ operators and commutes with $\hat{\boldsymbol{e}}_t$, $\hat{\boldsymbol{i}}$, $\hat{\boldsymbol{j}}$, $\hat{\boldsymbol{k}}$ while $\hat{R}_t$ anti-commutes with $\hat{\boldsymbol{e}}_t$, $\hat{\boldsymbol{e}}_{rt}$ and commutes with $\hat{\boldsymbol{e}}_r$, $\hat{\boldsymbol{i}}$, $\hat{\boldsymbol{j}}$, $\hat{\boldsymbol{k}}$. Moreover operators $\hat{R}_r$ and $\hat{R}_t$ commute with each other. Also we can introduce the operator of space-time inversion

$$\hat{R}_{rt} = \hat{R}_r \hat{R}_t \,, \tag{61}$$

which has the following properties:

$$\begin{aligned} \hat{R}_{rt}\psi_0 &= (A_0 - B_0\boldsymbol{e_r} - C_0\boldsymbol{e_t} + D_0\boldsymbol{e_{rt}})\,, \\ \hat{R}_{rt}\psi_1 &= (A_1 - B_1\boldsymbol{e_r} - C_1\boldsymbol{e_t} + D_1\boldsymbol{e_{rt}})\,, \\ \hat{R}_{rt}\psi_2 &= (A_2 - B_2\boldsymbol{e_r} - C_2\boldsymbol{e_t} + D_2\boldsymbol{e_{rt}})\,, \\ \hat{R}_{rt}\psi_3 &= (A_3 - B_3\boldsymbol{e_r} - C_3\boldsymbol{e_t} + D_3\boldsymbol{e_{rt}})\,. \end{aligned} \tag{62}$$

Above considered operators of spatial and time inversion allow us to construct the family of first-order equations. Let us turn to sedeonic equation (39)

$$\left(\frac{1}{c}\frac{\partial}{\partial t} - \hat{\boldsymbol{\alpha}}\vec{\nabla}\right)\left(\frac{1}{c}\frac{\partial}{\partial t} + \hat{\boldsymbol{\alpha}}\vec{\nabla}\right)\tilde{\psi} = -\frac{m^2c^2}{\hbar^2}\tilde{\psi} \tag{63}$$

corresponding to the Einstein relation for energy and momentum. Let $\hat{\boldsymbol{\alpha}} = \hat{\boldsymbol{e}}_r$ for definiteness. Then (63) can be rewritten as

$$\left(\frac{1}{c}\frac{\partial}{\partial t} - \hat{\boldsymbol{e}}_{\boldsymbol{r}}\vec{\nabla}\right)\left(\frac{1}{c}\frac{\partial}{\partial t} + \hat{\boldsymbol{e}}_{\boldsymbol{r}}\vec{\nabla}\right)\tilde{\psi} = -\frac{m^2c^2}{\hbar^2}\tilde{\psi}\,. \tag{64}$$

In this equation we can formally denote the result of action of one operator on function $\tilde{\psi}$ as some new sedeonic function $\tilde{\chi}$:

$$\left(\frac{1}{c}\frac{\partial}{\partial t} + \hat{\boldsymbol{e}}_{\boldsymbol{r}}\vec{\nabla}\right)\tilde{\psi} = -\frac{mc}{\hbar}\tilde{\chi}\,. \tag{65}$$

Then the second-order equation (64) is equivalent to the system of two first-order equations:

$$\begin{cases} \left(\frac{1}{c}\frac{\partial}{\partial t} + \hat{\boldsymbol{e}}_{\boldsymbol{r}}\vec{\nabla}\right)\tilde{\psi} = -\frac{mc}{\hbar}\tilde{\chi}, \\ \left(\frac{1}{c}\frac{\partial}{\partial t} - \hat{\boldsymbol{e}}_{\boldsymbol{r}}\vec{\nabla}\right)\tilde{\chi} = \frac{mc}{\hbar}\tilde{\psi}. \end{cases} \tag{66}$$

Acting on the second equation of (66) by the operator of spatial inversion $\hat{R}_r$ we get

$$\begin{cases} \left(\frac{1}{c}\frac{\partial}{\partial t} + \hat{\boldsymbol{e}}_{\boldsymbol{r}}\vec{\nabla}\right)\tilde{\psi} = \frac{mc}{\hbar}(-\tilde{\chi}), \\ \left(\frac{1}{c}\frac{\partial}{\partial t} + \hat{\boldsymbol{e}}_{\boldsymbol{r}}\vec{\nabla}\right)\hat{R}_r\tilde{\chi} = \frac{mc}{\hbar}\hat{R}_r\tilde{\psi}. \end{cases} \tag{67}$$

On some conditions the equations of system (67) can be absolutely equivalent. For that functions $\tilde{\chi}$ and $\tilde{\psi}$ should satisfy the following relations:

$$\begin{cases} \tilde{\psi} = \eta\hat{R}_r\tilde{\chi}, \\ -\tilde{\chi} = \eta\hat{R}_r\tilde{\psi}, \end{cases} \tag{68}$$

where $\eta$ is some constant. In particular for scalar $\eta$ we obtain $\eta = \pm\xi$. So if the wave function $\tilde{\psi}$ and accessory function $\tilde{\chi}$ satisfy the condition

$$\tilde{\chi} = \pm\xi\hat{R}_r\tilde{\psi}\,, \tag{69}$$

then the wave function $\tilde{\psi}$ satisfies the first-order equation. The sign in (69) can be chosen arbitrarily. If $\tilde{\chi} = +\xi \hat{R}_r \tilde{\psi}$ then the first-order equation has the following form:

$$\left( \frac{1}{c}\frac{\partial}{\partial t} + \hat{\boldsymbol{e}}_{\boldsymbol{r}} \vec{\nabla} + \xi \frac{mc}{\hbar} \hat{R}_r \right) \tilde{\psi} = 0 . \qquad (70)$$

Note that we can also act by $\hat{R}_r$ on the first equation of system (66). Then we get an equation with other sign before the gradient operator. Thus the sedeonic first-order equation can be written in four different forms:

$$\left( \frac{1}{c}\frac{\partial}{\partial t} + \hat{\boldsymbol{e}}_{\boldsymbol{r}} \vec{\nabla} + \xi \frac{mc}{\hbar} \hat{R}_r \right) \tilde{\psi} = 0 , \qquad (71)$$

$$\left( \frac{1}{c}\frac{\partial}{\partial t} + \hat{\boldsymbol{e}}_{\boldsymbol{r}} \vec{\nabla} - \xi \frac{mc}{\hbar} \hat{R}_r \right) \tilde{\psi} = 0 , \qquad (72)$$

$$\left( \frac{1}{c}\frac{\partial}{\partial t} - \hat{\boldsymbol{e}}_{\boldsymbol{r}} \vec{\nabla} + \xi \frac{mc}{\hbar} \hat{R}_r \right) \tilde{\psi} = 0 , \qquad (73)$$

$$\left( \frac{1}{c}\frac{\partial}{\partial t} - \hat{\boldsymbol{e}}_{\boldsymbol{r}} \vec{\nabla} - \xi \frac{mc}{\hbar} \hat{R}_r \right) \tilde{\psi} = 0 . \qquad (74)$$

In case of plane wave solutions all these equations give us the right dispersion relation

$$\left( E^2 - p^2 c^2 - m^2 c^4 \right)^8 = 0 , \qquad (75)$$

where $p^2 = p_x^2 + p_y^2 + p_z^2$. The roots of equation (75) $E = \pm\sqrt{p^2 c^2 + m^2 c^4}$ are eighthly degenerate.

In derivation of (71)-(74) we used anti-commutation of $\hat{R}_r$ and $\hat{\boldsymbol{e}}_{\boldsymbol{r}}$ operators. But operator $\hat{R}_{rt}$ has the same property. Therefore we can write the first-order equations also in the following form:

$$\left( \frac{1}{c}\frac{\partial}{\partial t} + \hat{\boldsymbol{e}}_{\boldsymbol{r}} \vec{\nabla} + \xi \frac{mc}{\hbar} \hat{R}_{rt} \right) \tilde{\psi} = 0 , \qquad (76)$$

$$\left( \frac{1}{c}\frac{\partial}{\partial t} + \hat{\boldsymbol{e}}_{\boldsymbol{r}} \vec{\nabla} - \xi \frac{mc}{\hbar} \hat{R}_{rt} \right) \tilde{\psi} = 0 , \qquad (77)$$

$$\left( \frac{1}{c}\frac{\partial}{\partial t} - \hat{\boldsymbol{e}}_{\boldsymbol{r}} \vec{\nabla} + \xi \frac{mc}{\hbar} \hat{R}_{rt} \right) \tilde{\psi} = 0 , \qquad (78)$$

$$\left( \frac{1}{c}\frac{\partial}{\partial t} - \hat{\boldsymbol{e}}_{\boldsymbol{r}} \vec{\nabla} - \xi \frac{mc}{\hbar} \hat{R}_{rt} \right) \tilde{\psi} = 0 . \qquad (79)$$

If we use in (63) $\hat{\alpha} = \hat{\boldsymbol{e}}_{\boldsymbol{t}}$ and choose operator $\hat{R}_t$ for separation, then we obtain the following set of first-order equations:

$$\left( \frac{1}{c}\frac{\partial}{\partial t} + \hat{\boldsymbol{e}}_{\boldsymbol{t}} \vec{\nabla} + \xi \frac{mc}{\hbar} \hat{R}_t \right) \tilde{\psi} = 0 , \qquad (80)$$

$$\left( \frac{1}{c}\frac{\partial}{\partial t} + \hat{\boldsymbol{e}}_{\boldsymbol{t}} \vec{\nabla} - \xi \frac{mc}{\hbar} \hat{R}_t \right) \tilde{\psi} = 0 , \qquad (81)$$

$$\left( \frac{1}{c}\frac{\partial}{\partial t} - \hat{\boldsymbol{e}}_{\boldsymbol{t}} \vec{\nabla} + \xi \frac{mc}{\hbar} \hat{R}_t \right) \tilde{\psi} = 0 , \qquad (82)$$

$$\left(\frac{1}{c}\frac{\partial}{\partial t}-\hat{\boldsymbol{e}}_t\vec{\nabla}-\xi\frac{mc}{\hbar}\hat{R}_t\right)\tilde{\psi}=0, \qquad (83)$$

which have different space-time properties. On the other hand if we choose operator $\hat{R}_{rt}$ for separation then the system of the first-order equation can be written in the following form:

$$\left(\frac{1}{c}\frac{\partial}{\partial t}+\hat{\boldsymbol{e}}_t\vec{\nabla}+\xi\frac{mc}{\hbar}\hat{R}_{rt}\right)\psi=0, \qquad (84)$$

$$\left(\frac{1}{c}\frac{\partial}{\partial t}+\hat{\boldsymbol{e}}_t\vec{\nabla}-\xi\frac{mc}{\hbar}\hat{R}_{rt}\right)\tilde{\psi}=0, \qquad (85)$$

$$\left(\frac{1}{c}\frac{\partial}{\partial t}-\hat{\boldsymbol{e}}_t\vec{\nabla}+\xi\frac{mc}{\hbar}\hat{R}_{rt}\right)\tilde{\psi}=0, \qquad (86)$$

$$\left(\frac{1}{c}\frac{\partial}{\partial t}-\hat{\boldsymbol{e}}_t\vec{\nabla}-\xi\frac{mc}{\hbar}\hat{R}_{rt}\right)\tilde{\psi}=0. \qquad (87)$$

Analogously for $\hat{\boldsymbol{\alpha}}=\hat{\boldsymbol{e}}_{rt}$ we can indicate two different systems of first-order equations. For $\hat{R}_r$ operator we get

$$\left(\frac{1}{c}\frac{\partial}{\partial t}+\hat{\boldsymbol{e}}_{rt}\vec{\nabla}+\xi\frac{mc}{\hbar}\hat{R}_r\right)\psi=0, \qquad (88)$$

$$\left(\frac{1}{c}\frac{\partial}{\partial t}+\hat{\boldsymbol{e}}_{rt}\vec{\nabla}-\xi\frac{mc}{\hbar}\hat{R}_r\right)\psi=0, \qquad (89)$$

$$\left(\frac{1}{c}\frac{\partial}{\partial t}-\hat{\boldsymbol{e}}_{rt}\vec{\nabla}+\xi\frac{mc}{\hbar}\hat{R}_r\right)\psi=0, \qquad (90)$$

$$\left(\frac{1}{c}\frac{\partial}{\partial t}-\hat{\boldsymbol{e}}_{rt}\vec{\nabla}-\xi\frac{mc}{\hbar}\hat{R}_r\right)\tilde{\psi}=0. \qquad (91)$$

If we use $\hat{R}_{rt}$ operator then we obtain the following equations:

$$\left(\frac{1}{c}\frac{\partial}{\partial t}+\hat{\boldsymbol{e}}_{rt}\vec{\nabla}+\xi\frac{mc}{\hbar}\hat{R}_t\right)\tilde{\psi}=0, \qquad (92)$$

$$\left(\frac{1}{c}\frac{\partial}{\partial t}+\hat{\boldsymbol{e}}_{rt}\vec{\nabla}-\xi\frac{mc}{\hbar}\hat{R}_t\right)\tilde{\psi}=0, \qquad (93)$$

$$\left(\frac{1}{c}\frac{\partial}{\partial t}-\hat{\boldsymbol{e}}_{rt}\vec{\nabla}+\xi\frac{mc}{\hbar}\hat{R}_t\right)\tilde{\psi}=0, \qquad (94)$$

$$\left(\frac{1}{c}\frac{\partial}{\partial t}-\hat{\boldsymbol{e}}_{rt}\vec{\nabla}-\xi\frac{mc}{\hbar}\hat{R}_t\right)\tilde{\psi}=0. \qquad (95)$$

It is clear that for any equation (71)-(74), (76)-(79), (80)-(83), (84)-(87), (88)-(91), (92)-(95) there is also the inverse procedure of obtaining the second-order equation analogous to the procedure used in the Dirac theory. For example, acting on the equation (71) by operator

$$\left(\frac{1}{c}\frac{\partial}{\partial t}-\hat{\boldsymbol{e}}_r\vec{\nabla}-\xi\frac{mc}{\hbar}\hat{R}_r\right), \qquad (96)$$

we get the following equation:

$$\left(\frac{1}{c}\frac{\partial}{\partial t}-\hat{\boldsymbol{e}}_r\vec{\nabla}-\xi\frac{mc}{\hbar}\hat{R}_r\right)\left(\frac{1}{c}\frac{\partial}{\partial t}+\hat{\boldsymbol{e}}_r\vec{\nabla}+\xi\frac{mc}{\hbar}\hat{R}_r\right)\tilde{\psi}=0. \qquad (97)$$

Multiplying the operators in the left part of (97) we can obtain the sedeonic second-order equation

$$\left(\frac{1}{c}\frac{\partial}{\partial t}-\hat{\boldsymbol{e}}_{\boldsymbol{r}}\vec{\nabla}\right)\left(\frac{1}{c}\frac{\partial}{\partial t}+\hat{\boldsymbol{e}}_{\boldsymbol{r}}\vec{\nabla}\right)\tilde{\psi}=-\frac{m^2c^2}{\hbar^2}\tilde{\psi}\ . \tag{98}$$

However, we specially emphasize though equation (98) coincides in form with the second-order equation (64), but the solutions of (98) should also satisfy the first-order equation (71) simultaneously. It essentially restricts the class of possible wave functions.

**7. Plane wave solutions of the first-order equations**

All first-order equations (71)-(74), (76)-(79), (80)-(83), (84)-(87), (88)-(91), (92)-(95) can be integrated into one sedeonic Dirac's-like equation

$$\left(\frac{1}{c}\frac{\partial}{\partial t}+\hat{\alpha}\,\vec{\nabla}+\xi\frac{mc}{\hbar}\hat{\rho}\right)\tilde{\psi}=0\,, \tag{99}$$

where operator $\hat{\alpha}$ takes any meaning from $\hat{\alpha}\in\{\pm\hat{\boldsymbol{e}}_{\boldsymbol{r}},\pm\hat{\boldsymbol{e}}_{\boldsymbol{t}},\pm\hat{\boldsymbol{e}}_{\boldsymbol{rt}}\}$ and operator $\hat{\rho}$ takes any meaning from $\hat{\rho}\in\left\{\pm\hat{R}_r,\pm\hat{R}_t,\pm\hat{R}_{rt}\right\}$, simultaneously $\hat{\alpha}$ and $\hat{\rho}$ should anticommute

$$\hat{\alpha}\hat{\rho}\tilde{\psi}=-\hat{\rho}\hat{\alpha}\tilde{\psi}\,. \tag{100}$$

The equation (99) enables the plane wave solution. Let us search the solution in the form

$$\tilde{\psi}=\tilde{\boldsymbol{W}}\ exp\left\{-\xi\left(Et-(\vec{p}\cdot\vec{r})\right)/\hbar\right\}, \tag{101}$$

where $\tilde{\boldsymbol{W}}$ is the wave amplitude, $\vec{p}$ is the absolute vector of momentum. Then the equation (99) is transformed to

$$\left(E-c\hat{\alpha}\,\vec{p}-mc^2\hat{\rho}\right)\tilde{\boldsymbol{W}}=0\,. \tag{102}$$

Representing $\tilde{\boldsymbol{W}}=\boldsymbol{W_0}+\vec{\boldsymbol{W}}$ we obtain the following equation:

$$E\boldsymbol{W_0}+E\vec{\boldsymbol{W}}-c\hat{\alpha}\,\vec{p}\,\boldsymbol{W_0}-c\hat{\alpha}\left(\vec{p}\cdot\vec{\boldsymbol{W}}\right)-c\hat{\alpha}\left[\vec{p}\times\vec{\boldsymbol{W}}\right]-mc^2\hat{\rho}\,\boldsymbol{W_0}-mc^2\hat{\rho}\,\vec{\boldsymbol{W}}=0\,. \tag{103}$$

Separating sedeon-scalar and sedeon-vector parts we get the following system:

$$\begin{cases}E\boldsymbol{W_0}-c\hat{\alpha}\left(\vec{p}\cdot\vec{\boldsymbol{W}}\right)-mc^2\hat{\rho}\,\boldsymbol{W_0}=0,\\ E\vec{\boldsymbol{W}}-c\hat{\alpha}\,\vec{p}\,\boldsymbol{W_0}-c\hat{\alpha}\left[\vec{p}\times\vec{\boldsymbol{W}}\right]-mc^2\hat{\rho}\,\vec{\boldsymbol{W}}=0.\end{cases} \tag{104}$$

Let momentum is directed along the Z axis, so $\vec{p}=p\boldsymbol{k}$, where $p$ is the momentum module. Then the system (104) can be transformed in the following way:

$$\begin{cases}E\boldsymbol{W}_\mathbf{0}-c\hat{\alpha}\,p\,\boldsymbol{W}_\mathbf{z}-mc^2\hat{\rho}\,\boldsymbol{W}_\mathbf{0}=0,\\ E\boldsymbol{W}_\mathbf{z}-c\hat{\alpha}\,p\,\boldsymbol{W}_\mathbf{0}-mc^2\hat{\rho}\,\boldsymbol{W}_\mathbf{z}=0,\\ E\boldsymbol{W}_\mathbf{x}+\xi\,c\hat{\alpha}\,p\,\boldsymbol{W}_\mathbf{y}-mc^2\hat{\rho}\,\boldsymbol{W}_\mathbf{x}=0,\\ E\boldsymbol{W}_\mathbf{y}-\xi\,c\hat{\alpha}\,p\,\boldsymbol{W}_\mathbf{x}-mc^2\hat{\rho}\,\boldsymbol{W}_\mathbf{y}=0.\end{cases} \tag{105}$$

Thus we obtain the following relations between components of the wave function:

$$\boldsymbol{W}_\mathbf{z}=\frac{\hat{\alpha}}{cp}\left(E-mc^2\hat{\rho}\ \right)\boldsymbol{W}_\mathbf{0}\,, \tag{106}$$

$$\boldsymbol{W}_{\mathbf{y}} = \frac{\xi\hat{\boldsymbol{\alpha}}}{cp}\left(E - mc^2\hat{\rho}\right)\boldsymbol{W}_{\mathbf{x}}, \tag{107}$$

or inversed relations

$$\boldsymbol{W}_{\mathbf{0}} = \frac{\hat{\boldsymbol{\alpha}}}{cp}\left(E - mc^2\hat{\rho}\right)\boldsymbol{W}_{\mathbf{z}}, \tag{108}$$

$$\boldsymbol{W}_{\mathbf{x}} = -\frac{\xi\hat{\boldsymbol{\alpha}}}{cp}\left(E - mc^2\hat{\rho}\right)\boldsymbol{W}_{\mathbf{y}}. \tag{109}$$

Taking into account relations (106) and (107) we get the amplitude of wave function in the following form:

$$\tilde{\boldsymbol{W}} = \left(1 + \frac{\hat{\boldsymbol{\alpha}}}{cp}\left(E - mc^2\hat{\rho}\right)\boldsymbol{k}\right)\boldsymbol{W}_{\mathbf{0}} + \left(\boldsymbol{i} + \frac{\xi\hat{\boldsymbol{\alpha}}}{cp}\left(E - mc^2\hat{\rho}\right)\boldsymbol{j}\right)\boldsymbol{W}_{\mathbf{x}}, \tag{110}$$

where $\boldsymbol{W}_{\mathbf{0}}$ and $\boldsymbol{W}_{\mathbf{x}}$ are arbitrary sedeonic complex constants. The expression (110) can be represented also in the following compact form:

$$\tilde{\boldsymbol{W}} = \left(1 + \frac{\hat{\boldsymbol{\alpha}}}{cp}\left(E - mc^2\hat{\rho}\right)\boldsymbol{k}\right)\left(\boldsymbol{W}_{\mathbf{0}} + \boldsymbol{W}_{\mathbf{x}}\boldsymbol{i}\right). \tag{111}$$

## 8. Sedeonic first-order equations for massless particles

Using the results of section 6 we can indicate three types of sedeonic first-order equations for massless particles (neutrinos), which are differed by transformation properties. The equations (71)-(74) and (76)-(79) lead us to the following equations for massless particles:

$$\left(\frac{1}{c}\frac{\partial}{\partial t} + \hat{\boldsymbol{e}}_r\vec{\nabla}\right)\tilde{\boldsymbol{\psi}} = 0, \tag{112}$$

$$\left(\frac{1}{c}\frac{\partial}{\partial t} - \hat{\boldsymbol{e}}_r\vec{\nabla}\right)\tilde{\boldsymbol{\psi}} = 0. \tag{113}$$

Analogously from equations (80)-(83) and (84)-(87) we obtain:

$$\left(\frac{1}{c}\frac{\partial}{\partial t} + \hat{\boldsymbol{e}}_t\vec{\nabla}\right)\tilde{\boldsymbol{\psi}} = 0, \tag{114}$$

$$\left(\frac{1}{c}\frac{\partial}{\partial t} - \hat{\boldsymbol{e}}_t\vec{\nabla}\right)\tilde{\boldsymbol{\psi}} = 0. \tag{115}$$

Finally from equations (88)-(91) and (92)-(95) we get the third pair of equations for massless particles

$$\left(\frac{1}{c}\frac{\partial}{\partial t} + \hat{\boldsymbol{e}}_{rt}\vec{\nabla}\right)\tilde{\boldsymbol{\psi}} = 0, \tag{116}$$

$$\left(\frac{1}{c}\frac{\partial}{\partial t} - \hat{\boldsymbol{e}}_{rt}\vec{\nabla}\right)\tilde{\boldsymbol{\psi}} = 0. \tag{117}$$

All first-order equations (112)-(117) can be integrated into one sedeonic equation

$$\left(\frac{1}{c}\frac{\partial}{\partial t} + \hat{\boldsymbol{\alpha}}\,\vec{\nabla}\right)\tilde{\boldsymbol{\psi}} = 0, \tag{118}$$

where operator $\hat{\alpha}$ takes any meaning from $\hat{\alpha} \in \{\pm\hat{\boldsymbol{e}}_r, \pm\hat{\boldsymbol{e}}_t, \pm\hat{\boldsymbol{e}}_{rt}\}$.

The equation (118) enables the plane wave solution. Let us search the solution in the form

$$\tilde{\psi} = \tilde{\boldsymbol{W}}\ exp\left\{-\xi\left(Et - \left(\vec{p}\cdot\vec{r}\right)\right)/\hbar\right\}, \tag{119}$$

where $\tilde{\boldsymbol{W}}$ is the wave amplitude, $\vec{p}$ is the absolute vector of momentum. Then equation (118) is transformed to

$$\left(E - c\hat{\alpha}\ \vec{p}\right)\tilde{\boldsymbol{W}} = 0. \tag{120}$$

The dispersion relation for the equation (118) has the form

$$E = \gamma_\nu cp, \tag{121}$$

where $p$ is the momentum module, $\gamma_\nu = +1$ for neutrino and $\gamma_\nu = -1$ for antineutrino.

Let momentum is directed along the Z axis, so $\vec{p} = p\boldsymbol{k}$. Then the amplitude of the wave function can be obtained directly from the expression (111) if we take the mass equal to zero:

$$\tilde{\boldsymbol{W}} = \left\{1 + \gamma_\nu\hat{\alpha}\boldsymbol{k}\right\}\left(\boldsymbol{W}_\mathbf{0} + \boldsymbol{W}_\mathbf{x}\boldsymbol{i}\right), \tag{122}$$

where $\boldsymbol{W}_\mathbf{0}$ and $\boldsymbol{W}_\mathbf{x}$ are arbitrary sedeonic complex constants. Then the generalized plane wave solution for the equation (118) can be written in the following form:

$$\tilde{\psi} = \left(1 + \gamma_\nu\hat{\alpha}\boldsymbol{k}\right)\left(\boldsymbol{W}_\mathbf{0} + \boldsymbol{W}_\mathbf{x}\boldsymbol{i}\right) exp\left\{\xi p\left(z - \gamma_\nu ct\right)/\hbar\right\}. \tag{123}$$

Concluding this section we would like to indicate one special sedeonic first-order equation for the massless particle corresponding to $\hat{\alpha}$ = ±1. In this case the equation (118) takes the form

$$\left(\frac{1}{c}\frac{\partial}{\partial t} + \vec{\nabla}\right)\tilde{\psi} = 0, \tag{124}$$

or adjoint form

$$\left(\frac{1}{c}\frac{\partial}{\partial t} - \vec{\nabla}\right)\tilde{\psi} = 0, \tag{125}$$

which do not depend on space-time inversion at all. We emphasize that this equations does not correspond any first-order equation for the massive particles from the system of (99), so it is the separate case. The equation (124) has the following plane wave solution for particle ($\nu$):

$$\tilde{\psi}_\nu = \left(1 + \boldsymbol{k}\right)\left(\boldsymbol{W}_\mathbf{0} + \boldsymbol{W}_\mathbf{x}\boldsymbol{i}\right) exp\left\{\xi p\left(z - ct\right)/\hbar\right\} \tag{126}$$

and for the antiparticle ($\bar{\nu}$):

$$\tilde{\psi}_{\bar{\nu}} = \left(1 - \boldsymbol{k}\right)\left(\boldsymbol{W}_\mathbf{0} + \boldsymbol{W}_\mathbf{x}\boldsymbol{i}\right) exp\left\{\xi p\left(z + ct\right)/\hbar\right\}. \tag{127}$$

Consequently the equation (124) describes simultaneously the particle and the antiparticle. It is clearly seen from (126) and (127) that wave functions of massless particle and antiparticle are the eigenfunctions of the operator $\hat{\boldsymbol{k}}$ (see section 5). At that the wave function of particle corresponds to the state with the eigenvalue $\lambda = +1$ and the wave function of antiparticle corresponds to the state with the eigenvalue $\lambda = -1$. So the expressions (126) and (127) describe polarized particles. Contrary to (124) the equation (125) describes the particle in the state with the eigenvalue $\lambda = -1$ and antiparticle in the state with the eigenvalue $\lambda = +1$.

## 9. Sedeonic equations for quantum fields

On the basis of sedeonic wave functions we can define some quantum fields, which satisfy the first-order equations analogous to the Maxwell equations. Indeed, by means of operators $\hat{R}_r$, $\hat{R}_t$, and $\hat{R}_{rt}$, which anticommute with corresponding operators $\boldsymbol{e_r}\vec{\nabla}$, $\boldsymbol{e_t}\vec{\nabla}$, and $\boldsymbol{e_{rt}}\vec{\nabla}$, the equation (63) can be represented as octonic product of two operators. Using operators $\hat{\boldsymbol{\alpha}}$ and $\hat{\rho}$ we can write this equation in the following form:

$$\left(\frac{1}{c}\frac{\partial}{\partial t} - \hat{\boldsymbol{\alpha}}\,\vec{\nabla} - \xi\frac{mc}{\hbar}\hat{\rho}\right)\left(\frac{1}{c}\frac{\partial}{\partial t} + \hat{\boldsymbol{\alpha}}\,\vec{\nabla} + \xi\frac{mc}{\hbar}\hat{\rho}\right)\tilde{\psi} = 0\,. \tag{128}$$

Let us consider the sequential action of operators in (128). After the action of the first operator we obtain

$$\left(\frac{1}{c}\frac{\partial}{\partial t} + \hat{\boldsymbol{\alpha}}\,\vec{\nabla} + \xi\frac{mc}{\hbar}\hat{\rho}\right)\tilde{\psi} = \frac{1}{c}\frac{\partial \psi_0}{\partial t} + \frac{1}{c}\frac{\partial \vec{\psi}}{\partial t} +$$

$$+\hat{\boldsymbol{\alpha}}\,\vec{\nabla}\psi_0 + \hat{\boldsymbol{\alpha}}\left(\vec{\nabla}\cdot\vec{\psi}\right) + \hat{\boldsymbol{\alpha}}\left[\vec{\nabla}\times\vec{\psi}\right] + \xi\frac{mc}{\hbar}\hat{\rho}\psi_0 + \xi\frac{mc}{\hbar}\hat{\rho}\vec{\psi}\,. \tag{129}$$

Let us introduce the complex quantum fields. We will indicate these fields by index $\psi$:

$$\boldsymbol{\varepsilon}_\psi = \frac{1}{c}\frac{\partial \psi_0}{\partial t} + \hat{\boldsymbol{\alpha}}\left(\vec{\nabla}\cdot\vec{\psi}\right) + \xi\frac{mc}{\hbar}\hat{\rho}\psi_0\,, \tag{130}$$

$$\vec{\boldsymbol{E}}_\psi = -\hat{\boldsymbol{\alpha}}\,\vec{\nabla}\psi_0 - \frac{1}{c}\frac{\partial \vec{\psi}}{\partial t} - \xi\frac{mc}{\hbar}\hat{\rho}\vec{\psi} - \hat{\boldsymbol{\alpha}}\left[\vec{\nabla}\times\vec{\psi}\right]. \tag{131}$$

Here $\boldsymbol{\varepsilon}_\psi$ is a sedeon-scalar field and $\vec{\boldsymbol{E}}_\psi$ is sedeon-vector field. Using field's definition the expression (129) can be rewritten in the form

$$\left(\frac{1}{c}\frac{\partial}{\partial t} + \hat{\boldsymbol{\alpha}}\,\vec{\nabla} + \xi\frac{mc}{\hbar}\hat{\rho}\right)\tilde{\psi} = \boldsymbol{\varepsilon}_\psi - \vec{\boldsymbol{E}}_\psi\,. \tag{132}$$

Then equation (128) can be rewritten as

$$\left(\frac{1}{c}\frac{\partial}{\partial t} - \hat{\boldsymbol{\alpha}}\,\vec{\nabla} - \xi\frac{mc}{\hbar}\hat{\rho}\right)\left(\boldsymbol{\varepsilon}_\psi - \vec{\boldsymbol{E}}_\psi\right) = 0\,. \tag{133}$$

Performing sedeonic multiplication and separating sedeon-scalar and sedeon-vector parts we obtain the system of the first-order equations for quantum fields:

$$\begin{cases} \left(\hat{\boldsymbol{\alpha}}\vec{\nabla}\cdot\vec{\boldsymbol{E}}_\psi\right) = -\dfrac{1}{c}\dfrac{\partial \boldsymbol{\varepsilon}_\psi}{\partial t} + \xi\dfrac{mc}{\hbar}\hat{\rho}\boldsymbol{\varepsilon}_\psi\,, \\ \left[\hat{\boldsymbol{\alpha}}\vec{\nabla}\times\vec{\boldsymbol{E}}_\psi\right] = \dfrac{1}{c}\dfrac{\partial \vec{\boldsymbol{E}}_\psi}{\partial t} + \hat{\boldsymbol{\alpha}}\,\vec{\nabla}\boldsymbol{\varepsilon}_\psi - \xi\dfrac{mc}{\hbar}\hat{\rho}\vec{\boldsymbol{E}}_\psi\,. \end{cases} \tag{134}$$

This system is absolutely equivalent to the equation (128).

Following the ideas of Refs. 26 and 28 we can generalize the system (134) for the case of a particle in an external electromagnetic field. Using substitution (42) we can write the equation (128) in the following form:

$$\left(\frac{1}{c}\frac{\partial}{\partial t} + \frac{\xi e}{\hbar c}\Phi - \hat{\boldsymbol{\alpha}}\,\vec{\nabla} + \frac{\xi e}{\hbar c}\hat{\boldsymbol{\alpha}}\vec{A} - \xi\frac{mc}{\hbar}\hat{\rho}\right)\left(\frac{1}{c}\frac{\partial}{\partial t} + \frac{\xi e}{\hbar c}\Phi + \hat{\boldsymbol{\alpha}}\,\vec{\nabla} - \frac{\xi e}{\hbar c}\hat{\boldsymbol{\alpha}}\vec{A} + \xi\frac{mc}{\hbar}\hat{\rho}\right)\tilde{\psi} = 0\,. \tag{135}$$

This equation enables the introduction of sedeon-scalar $\varepsilon_\psi$ and sedeon-vector $\vec{\boldsymbol{E}}_\psi$ quantum fields

$$\left(\frac{1}{c}\frac{\partial}{\partial t}+\frac{\xi e}{\hbar c}\Phi+\hat{\boldsymbol{\alpha}}\,\vec{\nabla}-\frac{\xi e}{\hbar c}\hat{\boldsymbol{\alpha}}\vec{A}+\xi\frac{mc}{\hbar}\hat{\rho}\right)\tilde{\boldsymbol{\psi}}=\varepsilon_\psi-\vec{\boldsymbol{E}}_\psi\,, \tag{136}$$

or in expanded form

$$\varepsilon_\psi=\frac{1}{c}\frac{\partial\psi_0}{\partial t}+\hat{\boldsymbol{\alpha}}\left(\vec{\nabla}\cdot\vec{\boldsymbol{\psi}}\right)+\xi\frac{mc}{\hbar}\hat{\rho}\psi_0+\frac{\xi e}{\hbar c}\Phi\psi_0-\frac{\xi e}{\hbar c}\hat{\boldsymbol{\alpha}}\left(\vec{A}\cdot\vec{\boldsymbol{\psi}}\right), \tag{137}$$

$$\vec{\boldsymbol{E}}_\psi=-\hat{\boldsymbol{\alpha}}\,\vec{\nabla}\psi_0-\frac{1}{c}\frac{\partial\vec{\boldsymbol{\psi}}}{\partial t}-\xi\frac{mc}{\hbar}\hat{\rho}\vec{\boldsymbol{\psi}}-\hat{\boldsymbol{\alpha}}\left[\vec{\nabla}\times\vec{\boldsymbol{\psi}}\right]-\frac{\xi e}{\hbar c}\Phi\,\vec{\boldsymbol{\psi}}+\frac{\xi e}{\hbar c}\hat{\boldsymbol{\alpha}}\vec{A}\psi_0+\frac{\xi e}{\hbar c}\hat{\boldsymbol{\alpha}}\left[\vec{A}\times\vec{\boldsymbol{\psi}}\right]. \tag{138}$$

Then using the field's definition we can write the equation (135) as

$$\left(\frac{1}{c}\frac{\partial}{\partial t}+\frac{\xi e}{\hbar c}\Phi-\hat{\boldsymbol{\alpha}}\,\vec{\nabla}+\frac{\xi e}{\hbar c}\hat{\boldsymbol{\alpha}}\vec{A}-\xi\frac{mc}{\hbar}\hat{\rho}\right)\left(\varepsilon_\psi-\vec{\boldsymbol{E}}_\psi\right)=0\,. \tag{139}$$

This equation leads us to the following Maxwell's like first-order sedeonic equations for quantum fields:

$$\begin{cases}\left(\hat{\boldsymbol{\alpha}}\vec{\nabla}\cdot\vec{\boldsymbol{E}}_\psi\right)=-\frac{1}{c}\frac{\partial\varepsilon_\psi}{\partial t}-\frac{\xi e}{\hbar c}\Phi\varepsilon_\psi+\frac{\xi e}{\hbar c}\hat{\boldsymbol{\alpha}}\left(\vec{A}\cdot\vec{\boldsymbol{E}}_\psi\right)+\xi\frac{mc}{\hbar}\hat{\rho}\varepsilon_\psi\,,\\ \left[\hat{\boldsymbol{\alpha}}\vec{\nabla}\times\vec{\boldsymbol{E}}_\psi\right]=\frac{1}{c}\frac{\partial\vec{\boldsymbol{E}}_\psi}{\partial t}+\hat{\boldsymbol{\alpha}}\,\vec{\nabla}\varepsilon_\psi+\frac{\xi e}{\hbar c}\Phi\vec{\boldsymbol{E}}_\psi-\frac{\xi e}{\hbar c}\hat{\boldsymbol{\alpha}}\vec{A}\varepsilon_\psi+\frac{\xi e}{\hbar c}\hat{\boldsymbol{\alpha}}\left[\vec{A}\times\vec{\boldsymbol{E}}_\psi\right]-\xi\frac{mc}{\hbar}\hat{\rho}\vec{\boldsymbol{E}}_\psi\,.\end{cases} \tag{140}$$

In conclusion to this section we note that in fact the generalized first order Dirac's-like equation (99) and corresponding equation for the particle in external electromagnetic field

$$\left(\frac{1}{c}\frac{\partial}{\partial t}+\frac{\xi e}{\hbar c}\Phi+\hat{\boldsymbol{\alpha}}\,\vec{\nabla}-\frac{\xi e}{\hbar c}\hat{\boldsymbol{\alpha}}\vec{A}+\xi\frac{mc}{\hbar}\hat{\rho}\right)\tilde{\boldsymbol{\psi}}=0 \tag{141}$$

describe particles, which do not have the quantum fields $\varepsilon_\psi$ and $\vec{\boldsymbol{E}}_\psi$ (see expressions (132) and (136)).

## 10. Conclusion

Thus in this paper we represented sixteen-component values "sedeons", generating associative noncommutative algebra. The sedeon is the complicated space-time object consisting of absolute scalar, space scalar, time scalar, space-time scalar, absolute vector, space vector, time vector, and space-time vector. All these values are differed with respect to spatial and time inversion.

On the basis of sedeonic algebra the generalized sedeonic equation of electrodynamics has been proposed. It was shown that there are several types of sedeonic wave equations for electromagnetic potentials, which are differed in transformational properties. However, it was demonstrated that the system of first-order equations (Maxwell equations) for electromagnetic fields can be formulated in the terms of absolute values.

We proposed a scheme for constructing relativistic quantum mechanics using sedeonic space-time operators and sedeonic wave functions. It was shown that the sedeonic second-order equation, corresponding to the Einstein relation between energy and momentum, correctly describes the interaction between spin and the electromagnetic field. It is established that the sedeonic wave function of a particle in the state with defined spin projection has the specific space-time structure in the form of a sedeonic oscillator with two spatial polarizations: longitudinal linear and transverse circular.

It was demonstrated that for the special class of wave functions the sedeonic second-order equation can be reduced to the single sedeonic first-order equation analogous to the Dirac equation. In dependence of space-time operators these equations describe three different kinds of particles (leptons) which are differed by space-time transformational properties. Besides we proposed three kinds of sedeonic first-order equations for massless particles (neutrinos) corresponding to the three types of leptonic equations. Moreover we indicated the special type of absolute equation for massless particle, which does not correspond any leptonic equation at all.

We showed that the sedeonic second-order wave equation can be reformulated in the form of the system of the first-order Maxwell's-like equations for the quantum fields. At the same time it was shown that the sedeonic Dirac's-like first-order equations describe particles, which do not have quantum fields.

**Acknowledgement**

The authors are very thankful to G.V. Mironova for kind assistance and moral support.

## APPENDIX A

*Table 3. The multiplication and commutation rules of sedeonic basis elements.*

| | $i$ | $j$ | $k$ | $i_r$ | $j_r$ | $k_r$ | $i_t$ | $j_t$ | $k_t$ | $i_{rt}$ | $j_{rt}$ | $k_{rt}$ | $e_r$ | $e_t$ | $e_{rt}$ |
|---|---|---|---|---|---|---|---|---|---|---|---|---|---|---|---|
| $i$ | 1 | $\xi k$ | $-\xi j$ | $e_r$ | $\xi k_r$ | $-\xi j_r$ | $e_t$ | $\xi k_t$ | $-\xi j_t$ | $e_{rt}$ | $\xi k_{rt}$ | $-\xi j_{rt}$ | $i_r$ | $i_t$ | $i_{rt}$ |
| $j$ | $-\xi k$ | 1 | $\xi i$ | $-\xi k_r$ | $e_r$ | $\xi i_r$ | $-\xi k_t$ | $e_t$ | $\xi i_t$ | $-\xi k_{rt}$ | $e_{rt}$ | $\xi i_{rt}$ | $j_r$ | $j_t$ | $j_{rt}$ |
| $k$ | $\xi j$ | $-\xi i$ | 1 | $\xi j_r$ | $-\xi i_r$ | $e_r$ | $\xi j_t$ | $-\xi i_t$ | $e_t$ | $\xi j_{rt}$ | $-\xi i_{rt}$ | $e_{rt}$ | $k_r$ | $k_t$ | $k_{rt}$ |
| $i_r$ | $e_r$ | $\xi k_r$ | $-\xi j_r$ | 1 | $\xi k$ | $-\xi j$ | $e_{rt}$ | $\xi k_{rt}$ | $-\xi j_{rt}$ | $e_t$ | $\xi k_t$ | $-\xi j_t$ | $i$ | $i_{rt}$ | $i_t$ |
| $j_r$ | $-\xi k_r$ | $e_r$ | $\xi i_r$ | $-\xi k$ | 1 | $\xi i$ | $-\xi k_{rt}$ | $e_{rt}$ | $\xi i_{rt}$ | $-\xi k_t$ | $e_t$ | $\xi i_t$ | $j$ | $j_{rt}$ | $j_t$ |
| $k_r$ | $\xi j_r$ | $-\xi i_r$ | $e_r$ | $\xi j$ | $-\xi i$ | 1 | $\xi j_{rt}$ | $-\xi i_{rt}$ | $e_{rt}$ | $\xi j_t$ | $-\xi i_t$ | $e_t$ | $k$ | $k_{rt}$ | $k_t$ |
| $i_t$ | $e_t$ | $\xi k_t$ | $-\xi j_t$ | $e_{rt}$ | $\xi k_{rt}$ | $-\xi j_{rt}$ | 1 | $\xi k$ | $-\xi j$ | $e_r$ | $\xi k_r$ | $-\xi j_r$ | $i_{rt}$ | $i$ | $i_r$ |
| $j_t$ | $-\xi k_t$ | $e_t$ | $\xi i_t$ | $-\xi k_{rt}$ | $e_{rt}$ | $\xi i_{rt}$ | $-\xi k$ | 1 | $\xi i$ | $-\xi k_r$ | $e_r$ | $\xi j_r$ | $j_{rt}$ | $j$ | $j_r$ |
| $k_t$ | $\xi j_t$ | $-\xi i_t$ | $e_t$ | $\xi j_{rt}$ | $-\xi i_{rt}$ | $e_{rt}$ | $\xi j$ | $-\xi i$ | 1 | $\xi j_r$ | $-\xi i_r$ | $e_r$ | $k_{rt}$ | $k$ | $k_r$ |
| $i_{rt}$ | $e_{rt}$ | $\xi k_{rt}$ | $-\xi j_{rt}$ | $e_t$ | $\xi k_t$ | $-\xi j_t$ | $e_r$ | $\xi k_r$ | $-\xi j_r$ | 1 | $\xi k$ | $-\xi j$ | $i_t$ | $i_r$ | $i$ |
| $j_{rt}$ | $-\xi k_{rt}$ | $e_{rt}$ | $\xi i_{rt}$ | $-\xi k_t$ | $e_t$ | $\xi i_t$ | $-\xi k_r$ | $e_r$ | $\xi i_r$ | $-\xi k$ | 1 | $\xi i$ | $j_t$ | $j_r$ | $j$ |
| $k_{rt}$ | $\xi j_{rt}$ | $-\xi i_{rt}$ | $e_{rt}$ | $\xi j_t$ | $-\xi i_t$ | $e_t$ | $\xi j_r$ | $-\xi i_r$ | $e_r$ | $\xi j$ | $-\xi i$ | 1 | $k_t$ | $k_r$ | $k$ |
| $e_r$ | $i_r$ | $j_r$ | $k_r$ | $i$ | $j$ | $k$ | $i_{rt}$ | $j_{rt}$ | $k_{rt}$ | $i_t$ | $j_t$ | $k_t$ | 1 | $e_{rt}$ | $e_t$ |
| $e_t$ | $i_t$ | $j_t$ | $k_t$ | $i_{rt}$ | $j_{rt}$ | $k_{rt}$ | $i$ | $j$ | $k$ | $i_r$ | $j_r$ | $k_r$ | $e_{rt}$ | 1 | $e_r$ |
| $e_{rt}$ | $i_{rt}$ | $j_{rt}$ | $k_{rt}$ | $i_t$ | $j_t$ | $k_t$ | $i_r$ | $j_r$ | $k_r$ | $i$ | $j$ | $k$ | $e_t$ | $e_r$ | 1 |

## References

1. L.D.Landau, E.M.Lifshits - Quantum Mechanics: Non-relativistic Theory (3rd ed.). London: Pergamon. Vol. 3 of the Course of Theoretical Physics. (1977).
2. V.B.Berestetskii, E.M.Lifshits, L.P.Pitaevskii - Quantum electrodynamics (2nd ed.). London: Pergamon. Vol. 4 of the Course of Theoretical Physics. (1982).
3. W. Pauli - Zur Quantenmechanik des magnetischen Elektrons, Zeitschrift für Physik, **43**(9-10), 601-623 (1927).
4. P.A.M.Dirac – The quantum theory of the electron, Proc. Roy. Soc. Lon. A., **117**, 610-624 (1928).
5. S.L.Adler – Quaternionic Quantum Mechanics and Quantum Fields, New York: Oxford University Press, 1995.
6. S.L.Adler – Time-dependent perturbation theory for quaternionic quantum mechanics, with application to CP nonconservation in K-meson decays, Phys. Rev. D, **34**(6), 1871-1877 (1986).
7. S.L.Adler – Quaternionic quantum field theory, Phys. Rev. Lett., **55**(8), 783-786 (1985).
8. S.L.Adler – Scattering and decay theory for quaternionic quantum mechanics, and the structure of induced T nonconservation, Phys. Rev. D, **37**(12), 3654-3662 (1988).
9. A.J.Davies, B.H.J.McKellar – Nonrelativistic quaternionic quantum mechanics in one dimension, Phys. Rev. A, **40**(8), 4209-4214 (1989).
10. A.J.Davies – Quaternionic Dirac equation, Phys. Rev. D, **41**(8), 2628-2630 (1990).
11. S.De Leo, P.Rotelli – Quaternion scalar field, Phys. Rev. D, **45**(2), 575-579 (1992).
12. S.De Leo - One Component Dirac Equation, (arXiv:hep-th/9508010v1 (1995)), Int. J. Mod. Phys. A, **11**, 3973-3986 (1996).
13. C.Schwartz – Relativistic quaternionic wave equation, Journal of Mathematical Physics, **47**, 122301 (2006).
14. L.Yu-Fen – Triality, biquaternion and vector representation of the Dirac equation, Advances in Applied Clifford Algebras, **12**(2), 109-124 (2002).
15. A.Gsponer, J.-P.Hurni - The physical heritage of sir W.R.Hamilton, arXiv:math-ph/0201058v2 (2002).
16. R.Penney – Octonions and Dirac equation, Amer. J. Phys., **36**, 871-873 (1968).
17. A.A.Bogush, Yu.A.Kurochkin – Cayley-Dickson procedure, relativistic wave equations and supersymmetric oscillators, Acta Applicandae Mathematicae, **50**, 121-129 (1998).
18. M. Gogberashvili - Octonionic version of Dirac equations, (arXiv:hep-ph/0505101 v2 (2005)), Int. J. Mod. Phys. A, **21**(17), 3513-3523 (2006).
19. S.De Leo, K.Abdel-Khalek – Octonionic Dirac equation, (arXiv:hep-th/9609033v1 (1996)), Prog. Theor. Phys., **96**, 833-846 (1996).
20. F. Gursey, C. H. Tze, On the role of division, Jordan and related algebras in particle physics (World Scientific, Singapore, 1996).
21. D.Hestenes – Observables, operators, and complex numbers in the Dirac theory, Journal of Mathematical Physics, **16**, 556-572 (1975).

22. D.Hestenes – Clifford algebra and the interpretation of quantum mechanics. In: Clifford algebra and their applications in mathematical physics. (Eds. J.S.R.Chisholm, A.K.Commons) Reidel, Dordrecht / Boston, 321-346 (1986).

23. W.M.Pezzaglia, A.W.Differ - A Clifford dyadic superfield from bilateral interaction of geometric multispin Dirac theory, arXiv:gr-qc/9311015v1 (1993).

24. V.L.Mironov, S.V.Mironov - Octonic electrodynamics, e-preprint arXiv:math-ph/0802.2435 (2008).

25. V.L.Mironov, S.V.Mironov – Octonic representation of electromagnetic field equations, Journal of Mathematical Physics, **50**, 012901 (2009).

26. V.L.Mironov, S.V.Mironov - Octonic relativistic quantum mechanics, e-preprint arXiv:math-ph/0803.0375 (2008).

27. V.L.Mironov, S.V.Mironov – Octonic second-order equations of relativistic quantum mechanics, Journal of Mathematical Physics, **50**, 012302 (2009).

28. V.L.Mironov, S.V.Mironov - Octonic first-order equations of relativistic quantum mechanics, International Journal of Modern Physics A (2009), (to be published).

29. K.Imaeda, M.Imaeda – Sedenions: algebra and analysis, Appl. Math. Comp., **115**, 77-88 (2000).

30. K.Carmody – Circular and hyperbolic quaternions, octonions, and sedenions, Appl. Math. Comput., **28**, 47-72 (1988).

31. K.Carmody – Circular and hyperbolic quaternions, octonions, and sedenions – further results, Appl. Math. Comput., **84**, 27-47 (1997).

32. J.Köplinger – Dirac equation on hyperbolic octonions, Appl. Math. Comput., **182**, 443-446 (2006).

33. W.P.Joyce – Dirac theory in spacetime algebra: I. The generalized bivector Dirac equation, J.Phys. A: Math. Gen., **34**, 1991-2005 (2001).

34. C.Cafaro, S.A. Ali – The spacetime algebra approach to massive classical electrodynamics with magnetic monopoles, Advances in Applied Clifford Algebras, **17**, 23-36 (2006).